\documentclass[doublespacing]{elsart}

\usepackage{times}
\usepackage{natbib}
\usepackage{amssymb}

\begin{document}

\begin{frontmatter}

\title{Kirchhoff rods with nonhomogeneous cross section}

\author{Alexandre F. da Fonseca and C. P. Malta\corauthref{cor1}}

\corauth[cor1]{To whom correspondence should be addressed (coraci@if.usp.br)}

\address{Instituto de F\'{\i}sica, Universidade de S\~ao Paulo,
USP\\ Rua do Mat\~{a}o Travessa R 187, CEP 05508-900, S\~ao Paulo,
Brazil}

\begin{abstract}

The Kirchhoff's theory for thin, inextensible, elastic rods with
nonhomogeneous cross section is studied. The Young's and shear moduli
of the rod are considered to vary radially, and it is shown that an
analytical solution for the constitutive relations can be obtained for
circular cross section and constant Poisson's ratio. We comment on
possible applications of our results.

\end{abstract}

\end{frontmatter}

Filamentous structures are of great interest in both academic and
industrial fields. From biological fibers and biopolymers to different
kinds of wires and cables, the study of statics and dynamics of these
systems has brought significant contribution to both the understanding
of the life and nature, and the development of new technological
devices.

The Kirchhoff rod model has been considered a good framework to study
both statics and dynamics of filaments in Biology
\citep{tamar1,tamar2,wolge,alain1,alain2,coleman,fonseca1} and in
Engineering \citep{sun,perkins}. In most cases, the rod is considered
to be completely homogeneous. But some nonhomogeneities along the rod
have been analyzed in the literature. Tridimensional conformations of
nonhomogeneous rods may present chaotic behavior
\citep{mielke,davies} and deviations from the helix pattern were
shown to occur in the case of a rod with periodic variation of its
Young's modulus \citep{fonseca2}. In order to investigate the
effects of stiffness nonhomogeneity, \citet{fonseca3} developed a
method for finding equilibrium solutions of the static Kirchhoff
equations for rods subjected to given boundary conditions. A geometric
study of rods with varying cross section radius was performed by
\citet{fonseca4}. In these examples, the rod nonhomogeneities are
function only of the arc-length along the rod axis (within a section
transversal to the axis the system is homogeneous).

Here we shall deal with nonhomogeneities that are function only of the
distance to the rod axis.  In this case, points lying at a given
distance from the rod axis will have the same mechanical properties
even if they belong to different cross sections. The elasticity
problem is defined as the specification of the so-called {\it state of
stress}. It requires the knowledge of the stress at every point of the
body \citep{love}. The work by \citet{zhang} and by \citet{chen} are
examples of application of Elasticity theory to nonhomogeneous
cylindrical rods. \citet{zhang} have obtained an exact solution for
the stress of a radially nonhomogeneous hollow circular cylinder with
exponential radial variation of the Young's modulus and constant
Poisson's ratio. Here, we address the problem of obtaining an
analytical solution for the stress of a nonhomogeneous rod within the
approximations of the Kirchhoff's theory for thin rods. We shall show
that an analytical solution can be obtained for a rod of circular
cross section presenting any kind of radial variation of the Young's
modulus, but constant Poisson's ratio. Since the Kirchhoff's rod
theory is largely used in modeling long, thin and inextensible elastic
rods, our solution can be of interest in a large range of
applications.

Examples of real systems presenting nonhomogeneous cross section are
coaxial cables \citep{tang}, coated optical fibers \citep{li} and
the double stranded DNA molecule \citep{livroDNA}. The process of
coating structures, where thin layers of a given material are
deposited on a given body, is largely used in industries as, for
example, the process called nitriding that improves the so-called {\it
tribo-mechanical} properties of engineering components
\citep{sylvio}.  Coating processes are also used in basic
research. For example, \citet{salvadori,salva2}
measured the Young's modulus of gold thin films deposited in Atomic
Force Microscopy (AFM) cantilevers. They found that the Young's
modulus of the gold thin films is about $12\%$ smaller than its bulk
elastic modulus. This information is interesting for the analysis we
shall present below.

The results of the Kirchhoff model were recently derived by
\citet{muller} through the rigorous method of $\Gamma$-convergence.
They showed that the nonlinear bending-torsion theory for inextensible
rods arises as the $\Gamma$-limit of three-di\-men\-si\-onal nonlinear
theory of elasticity. Nevertheless, we shall follow the more simple
derivation of the Kirchhoff equations due to \citet{dill} since our
final solution can be directly obtained using this approach. Our aim
is to obtain an analytical solution for the so-called {\it
constitutive relations} that relate the components of the moment to
the components of a vector representing the deformations of the rod.

\citet{dill} presented the derivation of the Kirchhoff equations
from the classical conservation laws of linear and angular momentum
for a three-di\-men\-si\-onal body with a surface area $A$ enclosing a
volume $V$:
\begin{equation}
\label{cl1} 
\begin{array}{l}
\int_{A}\mathbf{p}_{n}dS+\int_{V}\mathbf{f}dV=
\int_{V}\rho\ddot{\mathbf{X}}dV \; , \\
\int_{A}(\mathbf{X}\times\mathbf{p}_{n})dS+
\int_{V}(\mathbf{X}\times\mathbf{f})dV=
\int_{V}\rho(\ddot{\mathbf{X}}\times\mathbf{X})dV \; , 
\end{array}
\end{equation}
where $\mathbf{p}_{n}$ is the contact force per unit area exerted on
the oriented surface element $d\mathbf{S}=\mathbf{n}dS$, $\rho$ is the
mass density and $\mathbf{X}$ is the position with respect to a fixed
origin (a dot indicates time derivative). External forces per unit
volume acting on the body are represented by $\mathbf{f}$. We shall
drop the derivatives with respect to time since dynamical problems
will not be considered here.

The axis of the rod is defined as a smooth curve $\mathbf{x}$, in the
3D space, parametrized by the arc-length $s$:
$\mathbf{x}=\mathbf{x}(s)$. A {\it director basis} is defined at each
point of the curve, with $\mathbf{d}_3$ chosen as the tangent vector,
$\mathbf{d}_{3}=\mathbf{x}'$ (the prime denotes differentiation with
respect to $s$). The orthonormal vectors, $\mathbf{d}_1$ and
$\mathbf{d}_2$, lie in the plane normal to $\mathbf{d}_3$. We choose
these vectors in such a way that
$\mathbf{d}_1$,$\mathbf{d}_2$,$\mathbf{d}_3$ form a right-handed
orthonormal basis at each point of the rod axis. The space variation
of the director basis along the curve $\mathbf{x}$ is controlled by
the {\it twist equation}:
$\mathbf{d}_{i}'=\mathbf{k}\times\mathbf{d}_{i}$. The components in
the director basis, $k_1$ and $k_2$, of the {\it twist vector},
$\mathbf{k}$, are the components of the curvature of the rod, and
$k_3$ is the rod twist density.

In the Kirchhoff's theory the rod is seen as an assembly of short
segments. Each segment is loaded by contact forces from the adjacent
segments. The equations (\ref{cl1}) are applied to each segment in
order to obtain a one dimensional set of differential equations for
the static and dynamics of the rod. Each segment of infinitesimal
thickness will be referred to as a ``cross section''. The theory assumes
that the cross section radius, at all points, is much smaller than
both the total length and the curvature of the rod. Since equation
(\ref{cl1}) involves integration over the volume $V$ of the body or
over the area $A$ enclosing the volume $V$ of the body, the
nonhomogeneities in the cross section do not influence the main
Kirchhoff equations. Only the constitutive relations may depend on the
nonhomogeneities in the cross sections. The general equation relating
 stress and strain for a homogeneous isotropic material is
\citep{love,dill}
\begin{equation}
\label{stst} 
\tilde{S}=2\mu \tilde{E}+\lambda
(\mbox{Tr}\tilde{E})\tilde{\mathbf{1}} \; ,
\end{equation}
where the {\it tilde}\mbox{ } $\tilde{ }$\mbox{ } is used to denote a
tensor. $\tilde{S}$ is the stress tensor, $\tilde{E}$ is the strain
tensor and $\tilde{\mathbf{1}}$ is the unitary matrix. $\mu$ is the
shear modulus and $\lambda$ is one of the elastic constants of
Lam\'{e} \citep{love,dill}. The equation (\ref{stst}) is valid for
relatively small stresses where the linear theory of elasticity holds
true.

Our problem consists of obtaining the constitutive relations for a
rod made of a {\it nonhomogeneous isotropic material}. In order to
clarify the idea of {\it nonhomogeneity} and {\it isotropy} of the
material, we must remember that even though the length and the radius of
curvature of the rod have to be much bigger than the cross section
radius (for the validity of the Kirchhoff model), the rod is a
three-di\-men\-si\-onal body formed by elements of volume that are
considered to be a continuous isotropic medium. Nevertheless, some
of the elastic properties of the rod can vary from one element of
volume to another, as for example, in the radial direction. Elasticity
theory for nonhomogeneous isotropic media has been studied since 1960s
(see, for example \citep{rosto,plevako,chen2}).

Due to the fact that only rods with circular cross sections lead to an
analytical solution for the constitutive relations, so we shall
consider only radial variations for the elastic properties of the
rod. As in \citep{rosto,chen2}, the equation (\ref{stst}) remains
valid for nonhomogeneous $\mu$ and $\lambda$. Here, $\mu=\mu(r)$ and
$\lambda=\lambda(r)$, where $r$ is the distance of the point to the
axis of the rod.

The point where the axis intersects the plane of the cross section is
 the origin of a Cartesian basis lying in this plane. This
Cartesian basis can be defined by the two vectors $\mathbf{d}_1$ and
$\mathbf{d}_2$ of the director basis. The Young's modulus $E$ is
connected to the shear modulus $\mu$ through the Poisson's ratio, $\nu
$: $\nu +1=E/2\mu$.

The components $\epsilon_{ij}$ $(i,j=1,2,3)$ of the strain tensor,
$\tilde{E}$ (which is symmetric), within the Kirchhoff's theory, is
given by \citep{dill}:
\begin{equation}
\label{strain}
\begin{array}{l}
\epsilon_{\alpha \beta}=\frac{1}{2}
(\frac{\partial u_{\alpha}}{\partial X_{\beta}}+
 \frac{\partial u_{\beta}}{\partial X_{\alpha}}) \; , \\
\epsilon_{\alpha 3}=\frac{1}{2}
(\frac{\partial u_3}{\partial X_{\alpha}}+
(k_3-k_{3}^{(0)})X_{\beta}) \; , \\
\epsilon_{33}=(k_1-k_{1}^{(0)})X_2-(k_2-k_{2}^{(0)})X_1 \; ,
\end{array}
\end{equation}
where greek labels equal $1,2$, and $X_1$ and $X_2$ are the
components of the position vector of a material point of the rod in
the Cartesian basis $(\mathbf{d}_1 \ , \mathbf{d}_2)$. $k_{i}^{(0)}$
is the $i$-th component of the twist vector, in the director basis,
for the undeformed rod, also known as {\it intrinsic curvature}. 
$u_i$, $i=1,2,3$, are the components of the displacement of the
material point. The solution for $u_i=u_i(X_1,X_2)$ constitutes the
solution for the elasticity problem.

In order to find the solutions for  $u_i$ ($i=1,2,3$) we shall use
the local balance of momentum for the components of the stress tensor
\citep{dill}:
\begin{equation}
\label{local}
\sum_{\alpha=1}^{2}\frac{\partial S_{\alpha l}}{\partial X_{\alpha}}=0
\; , \; \; \; l=1,2,3 \; ,
\end{equation}
where $S_{ij}$ is the $ij$-th component of the stress tensor,
$\tilde{S}$. The boundary conditions are provided by the load
conditions on the rod lateral surface. In the Kirchhoff's theory, it is
given by
\begin{equation}
\label{boundary}
\sum_{\alpha=1}^{2}n_{\alpha}S_{\alpha l}=0 \; , \; \; \; l=1,2,3 \; ,
\end{equation}
where $n_1$ and $n_2$ are the components of the unit outward vector, 
normal to the boundary of the undeformed cross section. Therefore, the
equation (\ref{local}) must be solved for $u_i$ subjected to the
boundary conditions defined in (\ref{boundary}).

The equation (\ref{local}) can be separated and solved in two
sets. The first one includes all terms with the index $l=3$
(equation (\ref{local})):
\begin{equation}
\label{l_31} 
\sum_{\alpha=1}^{2}\frac{\partial S_{\alpha 3}}{\partial X_{\alpha}}=0 \; ,
\end{equation}
\begin{equation}
\label{l_32} 
S_{\alpha 3}=2\mu(r)\epsilon_{\alpha 3} \; ,
\end{equation}
\begin{equation}
\label{l_33} 
\epsilon_{\alpha 3}=\frac{1}{2}(\frac{\partial u_3}{\partial X_{\alpha}}+
(k_3-k_{3}^{(0)})X_{\beta}) \; .
\end{equation}
The boundary conditions for this set are:
\begin{equation}
\label{boundl_3} 
\sum_{\alpha=1}^{2}n_{\alpha}S_{\alpha 3}=0 \; .
\end{equation} 

The distance to the origin, $r$, is connected to $X_1$ and $X_2$
through the polar transformation $X_1=r\cos\theta$, $X_2=r\sin\theta$
($r=\sqrt{X_{1}^{2}+X_{2}^{2}}$). Substituting equations
(\ref{l_32}) and (\ref{l_33}) in equation (\ref{l_31}), and using the
following relation
\begin{equation}
\label{r_x} 
\frac{\partial r}{\partial X_{\alpha}}=\frac{X_{\alpha}}{r} \; ,
\end{equation}
 equation (\ref{l_31}) reads
\begin{equation}
\label{eq_u_3}
\frac{\partial }{\partial X_1}(\mu(r)\frac{\partial u_3}{\partial
X_1})+
\frac{\partial }{\partial X_2}(\mu(r)\frac{\partial u_3}{\partial
X_2})=0 \; . 
\end{equation}
This equation describes the torsion of a rod subjected to the boundary
condition given by equation (\ref{boundl_3}). The known solution is
\citep{love,dill}:
\begin{equation}
\label{sol_u_3} 
u_3=(k_3-k_{3}^{(0)})\varphi(X_1,X_2) \; ,
\end{equation}
where $\varphi(X_1,X_2)$ is known as {\it warping} \citep{dill} or
{\it torsion} \citep{love} function, and must satisfy:
\begin{equation}
\label{warping}
\mu(r)(\frac{\partial^{2} \varphi}{\partial X_{1}^{2}}+
\frac{\partial^{2} \varphi}{\partial X_{2}^{2}})+
\frac{1}{r}\frac{d\mu}{dr}(X_1\frac{\partial \varphi}{\partial X_1}+
X_2\frac{\partial \varphi}{\partial X_2})=0 \; . 
\end{equation}
The boundary condition (\ref{boundl_3}) becomes:
\begin{equation}
\label{bound_phi}
n_1(\frac{\partial \varphi}{\partial X_1}-X_2)+
n_2(\frac{\partial \varphi}{\partial X_2}+X_1)=0 \; ,
\end{equation}
and must be satisfied for all $X_1$ and $X_2$ such that
$\sqrt{X_{1}^{2}+X_{2}^{2}}=h$, $h$ being the cross section radius.

In the homogeneous case, the function $\varphi$ depends only on the
geometry of the cross section. In the nonhomogenoeus case, it also
depends on how the shear modulus $\mu$ varies with $r$. Nevertheless,
by inspection of the equation (\ref{bound_phi}) we see that the
solution for the homogeneous case with circular cross section
($\varphi(X_1,X_2)=0$) also satisfies the equation (\ref{warping}) and
the boundary condition (\ref{bound_phi}). Therefore, we consider here
only rods with circular cross section.

The second set consists of the remaining equations:
\begin{equation}
\label{l_12_1} 
\sum_{\alpha=1}^{2}\frac{\partial S_{\alpha \beta}}{\partial
X_{\alpha}}=0 \; ,
\end{equation}
\begin{equation}
\label{l_12_2} 
S_{\alpha \beta}=2\mu(r)\epsilon_{\alpha \beta}+
\lambda(r)(\sum_{m=1}^{3}\epsilon_{mm})\delta_{\alpha \beta} , \; ,
\end{equation}
\begin{equation}
\label{l_12_3} 
\epsilon_{\alpha \beta}=\frac{1}{2}(\frac{\partial u_{\alpha}}
{\partial X_{\beta}}+\frac{\partial u_{\beta}}{\partial X_{\alpha}}) \; ,
\end{equation}
where $\beta=1,2$ and $\delta_{\alpha \beta}$ is the Kronecker
delta. The boundary conditions for this set of equations are:
\begin{equation}
\label{bound_l12}
\begin{array}{l}
\sum_{\alpha=1}^{2}n_{\alpha}S_{\alpha 1}=0 \; , \\

\sum_{\alpha=1}^{2}n_{\alpha}S_{\alpha 2}=0 \; , 
\end{array}
\end{equation}
If the Poisson's ratio, $\nu$, is a constant within the cross section,
it is possible to show that the solution for $u_1$ and $u_2$ in
eqs. (\ref{l_12_1}-\ref{l_12_3}) have the same form of the homogeneous
case, even for $\mu=\mu(r)$ and $\lambda=\lambda(r)$, and the boundary
conditions (\ref{boundary}) are satisfied. Since the stresses
considered here are such that the rod remains inextensible, the
assumption of constant Poisson's ratio is reasonable in spite of
having varying Young's and shear moduli. In this case, the ratio of
Young's over shear moduli must satisfy $\frac{E(r)}{2\mu(r)}=$
Constant $=\nu+1$.

The explicit solutions for $u_1$ and $u_2$ are
\begin{equation}
\label{u_1_u_2}
\begin{array}{l}
u_1(X_1,X_2)=-\nu(k_1-k_{1}^{(0)})X_1X_2+
\frac{\nu}{2}(k_2-k_{2}^{(0)})(X_{1}^{2}-X_{2}^{2}) \; , \\
u_2(X_1,X_2)=\nu(k_2-k_{2}^{(0)})X_1X_2+
\frac{\nu}{2}(k_1-k_{1}^{(0)})(X_{1}^{2}-X_{2}^{2}) \; .
\end{array}
\end{equation}

Now, we can calculate the constitutive relations in terms of the
components of the twist vector. The definition of the total moment of
the cross section, $\mathbf{M}$, is
\begin{equation}
\label{totM}
\mathbf{M}=\int_{S}\mathbf{r}\times\mathbf{p}_S\,dS \; ,
\end{equation}
where $S$ is the area of the cross section, $\mathbf{r}$ is the
position vector in the plane of the cross section,  given by
\begin{equation}
\label{vec_r}
\mathbf{r}=X_1\mathbf{d}_1+X_2\mathbf{d}_2 \; ,
\end{equation}
and $\mathbf{p}_S$ is the contact force per unit area in the cross
section that, in terms of the stress tensor, is given by:
\begin{equation}
\label{ps}
\mathbf{p}_S=\mathbf{d}_3.\tilde{S}=S_{31}\mathbf{d}_1+
S_{32}\mathbf{d}_2+S_{33}\mathbf{d}_3 \; .
\end{equation}
Using the equations (\ref{vec_r}) and (\ref{ps}), and the fact that
$\mathbf{M}=M_1\mathbf{d}_1+M_2\mathbf{d}_2+M_3\mathbf{d}_3$, we
obtain:
\begin{equation}
\label{M1}
M_1=\int_SX_2S_{33}\,dS \; ,
\end{equation}
\begin{equation}
\label{M2}
M_2=\int_S-X_1S_{33}\,dS \; ,
\end{equation}
\begin{equation}
\label{M3}
M_3=\int_S(X_1S_{32}-X_1S_{31})\,dS \; .
\end{equation}

Finally, using the solutions for $u_i$ given by the equations
(\ref{u_1_u_2}), and by the solution $u_3=0$ (we are considering
circular cross section so that $\varphi(X_1,X_2)=0$) we can
obtain the components of the strain tensor using the equations
(\ref{l_33}) and (\ref{l_12_3}). Substituting the components of the
strain tensor in the equations (\ref{l_32}) and (\ref{l_12_2}) we can
obtain the expressions for the components $S_{31}$, $S_{32}$ and
$S_{33}$ of the stress tensor:
\begin{equation}
\label{S31}
S_{31}=-\mu(r)(k_3-k_{3}^{(0)})X_2 \; ,
\end{equation}
\begin{equation}
\label{S32}
S_{32}=\mu(r)(k_3-k_{3}^{(0)})X_1 \; ,
\end{equation}
\begin{equation}
\label{S33}
S_{33}=E(r)((k_1-k_{1}^{(0)})X_2-(k_2-k_{2}^{(0)})X_1) \; .
\end{equation}
The constitutive relations for the components of the total moment
of the cross section can be written in a final form as:
\begin{equation}
\label{fM1}
M_1=(k_1-k_{1}^{(0)})\pi\int_{0}^{h}E(r)r^{3}dr \; ,
\end{equation}
\begin{equation}
\label{fM2}
M_2=(k_2-k_{2}^{(0)})\pi\int_{0}^{h}E(r)r^{3}dr \; ,
\end{equation}
\begin{equation}
\label{fM3}
M_3=(k_3-k_{3}^{(0)})2\pi\int_{0}^{h}\mu(r)r^{3}dr \; ,
\end{equation}
where $h$ is the cross section radius. 

Note that there is no constraint on the variation of the Young's or
shear modulus with $r$ (provided that the Poisson's ratio is constant).

The constitutive relations given by equations (\ref{fM1}-\ref{fM3})
have the same form of the homogeneous case. It means that within the
approximations of the Kirchhoff's theory, the static and dynamics of a
rod are not affected by radial nonhomogeneities in its cross
sections. Our calculations constitute a demonstration that a thin
nonhomogeneous rod behaves like a homogeneous one if the deformations
have high radius of curvature as compared to the cross section radius.

The equations (\ref{fM1}-\ref{fM3}) can be used to compare the
rigidity of homogeneous and nonhomogeneous rods with the same cross
section radius. By comparing the expressions for $M_1$, $M_2$ and
$M_3$ of homogeneous and nonhomogeneous cases, we can derive
expressions for an {\it effective} Young's and shear moduli, $E_{ef}$
and $\mu_{ef}$, in terms of $E(r)$ and $\mu(r)$, respectively. By {\it
effective} we mean the values for Young's and shear moduli which
result in the same values for $M_i$ ($i=1,2,3$), if the rod was
homogeneous. They are given by:
\begin{equation}
\label{Eef}
E_{ef}=\frac{4}{h^{4}}\int_{0}^{h}E(r)r^{3}dr \; ,
\end{equation}
\begin{equation}
\label{muef}
\mu_{ef}=\frac{4}{h^{4}}\int_{0}^{h}\mu(r)r^{3}dr \; .
\end{equation}

Consider the simple example of a
rod like a coaxial cable, with  the Young's modulus  given by
\begin{equation}
\label{E0}
E(r)=\left\{ \begin{array}{l}
E_0 \; \; \mbox{for} \; \; 0<r<r_0 \; , \\
E_1 \; \; \mbox{for} \; \; r_0\leq r\leq h \; ,
\end{array} \right.
\end{equation}
so that $E_0$ and $E_1$ are the Young's moduli of the inner and outer parts of
the cross section defined by the regions from the origin to $r_0$ and
from $r_0$ to $h$, respectively. Substituting the equation (\ref{E0})
in the equation (\ref{Eef}), we obtain the following expression
for the effective Young's modulus:
\begin{equation}
\label{Eef2}
E_{ef}=E_0\left( \frac{r_0}{h} \right )^{4} + E_1 \left
(1-\frac{r_{0}^{4}}{h^{4}} \right ) \; .
\end{equation}

This simple expression can be used in various experimental set ups to
find out the Young's modulus of several systems. For example, consider
the experiment of \citet{salvadori,salva2} where the Young's modulus
of gold thin films deposited in Atomic Force Microscopy (AFM)
cantilevers were measured. We cannot apply our method directly to
their experiments because the cross sections of the cantilevers are
not circular, but our model provides another form of measuring the
Young's modulus of gold thin films. If a thin film of gold (or another
material) is made to grow on the cylindrical surface of a homogeneous
rod, which Young's modulus $E_0$ is known, by measuring the bending
coefficient of the coated rod (which is related to $E_{ef}$), the film
thickness, $h-r_0$,  and the total radius of the cross
section, $h$, we can use equation (\ref{Eef2}) to obtain the Young's
modulus, $E_1$, of the thin film. The only requirement is that the
dimensions of the rod must be in accordance with the approximations of the
Kirchhoff model.

Another interesting application is the study of the rigidity of
superficial layers of nitride and carbonitride compounds within steel
due to absortion of nitrogen by a process called nitriding
\citep{sylvio}. Our method can be used in two applications, in this 
case.

In the first application, as for the case of rods coated by thin
films, we can measure the Young's modulus of the substance of the
layer of nitride and carbonitride compounds by producing rods of steel
(which Young's modulus is known), submitting the rod to the nitriding
process and measuring the bending coefficient of the nitrided rod and
the thickness of the layer. As in the previous case, the equation
(\ref{Eef2}) gives  the Young's modulus of the layer.

The second application of our method to nitrided rods is an indirect
measurement of the thickness of the layer of nitride and carbonitride
compounds through the measurement of the Young's modulus of the nitrided
rod. In this case, the Young's modulus of the steel, and of the
substance that composes the layer, must be previously known. We can
obtain $r_0$ through equation (\ref{Eef2}).

In the Introduction, we mentioned that the Kirchhoff rod model has
been used to study the elastic behavior of long pieces of DNA
\citep{tamar1,tamar2,coleman}. Since it is known that DNA is not a
polymer with uniform cross section, our calculations guarantee that
those studies about elasticity of long DNA's are not inconsistent with
the use of the Kirchhoff rod model. Nevertheless, our approach could
be also used to obtain information about the stiffness of the parts of
the DNA. It is known that the most rigid part of the double helix DNA
is the phosphate backbone chain \citep{livroDNA}. The bases are
connected, in a base-pair, through hydrogen bonds that are weak
connections. Therefore, a complete base-pair in a DNA molecule is a
net situation approximately close to the cross section of a coaxial
cable. The inner part of DNA is formed by the bases that are connected
by weak hydrogen bonds and the outer part is formed by the strong
phosphate backbone chains. If, by means of molecular approaches, it is
possible to estimate the stiffness of the inner (outer) part of the
DNA molecule, then the stifness of the outer (inner) part could be
obtained, and the results could be checked by experimentally measured
stiffness of long DNAs (see, for example, \citep{busta}).

\section*{Acknowledgments}

The authors acknowledge the financial support from the Brazilian
agencies CAPES, CNPq and FAPESP. The authors would like to thank 
Dr. David Swigon for a critical reading of our manuscript.

\end{document}